\documentclass[]{aa}  

\usepackage{graphicx}
\begin{document}

\title{The GAPS Programme with HARPS-N@TNG}

\subtitle{XII: Characterization of the planetary system around HD\,108874 \thanks{Based on observations made with the Italian Telescopio Nazionale
Galileo (TNG) operated on the island of La Palma by the Fundaci\'on
Galileo Galilei of the INAF at the Spanish Observatorio del Roque
de los Muchachos of the IAC in the frame of the programme Global
Architecture of Planetary Systems (GAPS).}\fnmsep\thanks{Tab.\ref{RV}}}

\author{S. Benatti,\inst{1} 
 S. Desidera,\inst{1}  M. Damasso,\inst{2} L. Malavolta,\inst{3,1} A. F. Lanza,\inst{4} 
 K. Biazzo,\inst{4}  A.S. Bonomo,\inst{2}  R.U. Claudi,\inst{1}  F. Marzari,\inst{3} E. Poretti,\inst{5}
R. Gratton,\inst{1} G. Micela, \inst{6} I. Pagano, \inst{4} G. Piotto, \inst{3,1} A. Sozzetti,\inst{2}
C. Boccato,\inst{1} R. Cosentino,\inst{7} E. Covino, \inst{8} A. Maggio, \inst{6} E. Molinari,\inst{7} R. Smareglia,\inst{9} 
L. Affer,\inst{6} G. Andreuzzi,\inst{7} A. Bignamini,\inst{9} F. Borsa,\inst{5}  L. di Fabrizio,\inst{7}  M. Esposito,\inst{8}  
A. Martinez Fiorenzano,\inst{7} S. Messina,\inst{4}  P. Giacobbe,\inst{2} A. Harutyunyan,\inst{7} C. Knapic,\inst{9}  J. Maldonado,\inst{6}
 S. Masiero,\inst{1}  V. Nascimbeni,\inst{1}  M. Pedani,\inst{7}  M. Rainer,\inst{5}  G. Scandariato\inst{4} and R. Silvotti\inst{2}
}

\institute{INAF -- Osservatorio Astronomico di Padova,  Vicolo dell'Osservatorio 5, I-35122, Padova, Italy  \\
\email{serena.benatti@oapd.inaf.it}
\and  INAF -- Osservatorio Astrofisico di Torino, Via Osservatorio 20, I-10025, Pino Torinese, Italy 
\and Dipartimento di Fisica e Astronomia Galileo Galilei -- Universit\`a di Padova, Via Francesco Marzolo, 8, Padova PD  
\and INAF -- Osservatorio Astrofisico di Catania, Via S. Sofia 78, I-95123, Catania, Italy 
\and INAF -- Osservatorio Astronomico di Brera, Via E. Bianchi 46, I-23807 Merate (LC), Italy 
\and INAF -- Osservatorio Astronomico di Palermo, Piazza del Parlamento, 1, I-90134, Palermo, Italy 
\and Fundaci\'on Galileo Galilei - INAF, Rambla Jos\'e Ana Fernandez P\'erez 7, E-38712 Bre\~na Baja, TF - Spain 
\and INAF -- Osservatorio Astronomico di Capodimonte, Salita Moiariello 16, I-80131, Napoli, Italy 
\and INAF -- Osservatorio Astronomico di Trieste, via Tiepolo 11, I-34143 Trieste, Italy 
}
\date{Received ; accepted }

\abstract
{In order to understand  the observed physical and orbital diversity of extrasolar planetary systems, a full investigation of these objects and of their host stars is necessary.
Within this field, one of the main purposes of the GAPS observing project with HARPS-N@TNG is to provide a more detailed characterisation of already known systems. In this framework we monitored the star, hosting two giant planets, HD\,108874, with HARPS-N for three years in order to refine the orbits, to improve the dynamical study and to search for additional low-mass planets in close orbits. 
We subtracted the radial velocity (RV) signal due to the known outer planets, finding a clear modulation of 40.2 d period. We analysed the correlation between RV residuals and the activity indicators and modelled the magnetic activity with a dedicated code. 
Our analysis suggests that the 40.2 d periodicity is a signature of the rotation period of the star. A refined orbital solution is provided, revealing that the system is close to a mean motion resonance of about 9:2, in a stable configuration over 1 Gyr. Stable orbits for low-mass planets are limited to regions very close to the star or far from it. Our data exclude super-Earths with $M\sin i \gtrsim 5$ M$_{\oplus}$ within 0.4 AU and objects with M$\sin i \gtrsim 2$ M$_{\oplus}$ with orbital periods of a few days. Finally we put constraints on the habitable zone of the system, assuming the presence of an exomoon orbiting the inner giant planet. }

\titlerunning{Characterization of the planetary system around HD\,108874}
\authorrunning{S. Benatti et al.}

\maketitle
%

\section{Introduction} \label{intro}

One of the emerging scenarios of the exoplanets population, as revealed from the observations, is that planets are preferably organized in multiple systems (see e.g. \citealt{2011arXiv1109.2497M}; \citealt{2011ApJ...732L..24L}). Nevertheless, no evidence of a strict analog of our solar system has been found so far, but hundreds of systems with a large variety of architectures \citep{2014ApJ...790..146F}. The study of the planet distribution in multiple systems allows us to put constraints on theories for their complex formation, dynamics and evolution.
Radial velocity (RV) surveys can play a crucial role in this framework for several reasons. The availability of new generation instruments, able to push down the minimum-mass detection limit, allows nowadays to search for additional low-mass companions in systems already known to host giant planets. This provides a more complete view of their architecture as well as indications on the frequency of systems similar to the solar one. 
A full characterization of multiplanet systems requires great observational effort, as the number of components increase. In addition, several sources of astrophysical noise affect the RV (stellar oscillations, granulation and activity) and could impact on the measurements, either mimicking the presence of a planetary companion or tangling the interpretation of the signal \citep{2011A&A...527A..82D}.
Since August 2012 the GAPS (Global Architecture of Planetary Systems, see e.g. \citealt{2013A&A...554A..28C}) observing programme started its operations thanks to the high performances of the HARPS-N spectrograph \citep{2012SPIE.8446E..1VC}, mounted at the Italian telescope TNG in La Palma, Canary Islands. Within GAPS we studied the G9V star HD\,108874 in the framework of a dedicated sub-programme focused on the characterization of systems with already known planets. \cite{2003ApJ...582..455B} claimed the presence of a giant planet ($M \sin i$ = 1.71 $M_{\rm J}$) with a period of $397.5 \pm 4.0$ days (d), while \cite{2005ApJ...632..638V} found an additional companion on a wider orbit ($M \sin i = 1.02 \pm 0.3$ $M_{\rm J}$, $P_{\rm orb}$ = $1605.8 \pm 88.0$ d) through the analysis of $\sim$70 spectra from HIRES (Keck telescope). The presence of these planets was subsequently confirmed by other authors (\citealt{2009ApJ...693.1084W}, hereafter Wr09, and \citealt{2009ApJS..182...97W}, hereafter Wi09), while further studies \citep{2006ApJ...645..688G, 2007A&A...461..759L, 2010ApJ...715..803V} analysed the dynamics of this system in the presence of a 4:1 mean motion resonance between them. The planet HD\,108874 b is probably located in the habitable zone of its host star and this stimulated \cite{2007A&A...462.1165S} to investigate possible stable regions for Earth-like Trojan planets around it.
Here we present an updated analysis of the HD\,108874 system based on a three-year intensive monitoring with HARPS-N: we describe the observations, the data reduction and the derivation of stellar parameters in Sects. \ref{obs} and \ref{parameters}; we present our RV and stellar activity analysis in Sect. \ref{rva} and \ref{act}; a new orbital solution is shown in Sect. \ref{orbit}; the dynamical analysis is described in Sect. \ref{stab}; a discussion on the detection limits, the system architecture and its habitability is provided in Sect. \ref{discuss}.


\section{Observations and data reduction} \label{obs}
The GAPS observations of HD\,108874 with HARPS-N at TNG lasted three seasons, from December 2012 to July 2015.
During the first half of the observations the spectrograph was affected by a small defocus, evident in the time series of some parameters (FWHM and contrast of the Cross-Correlation Function, CCF) but negligible for others, like the bisector or the RVs, only causing a small increase in the estimated errors.
The problem was successfully fixed in March 2014. The total number of collected spectra is 94, spread over 82 nights.
The simultaneous Th-Ar calibration was used to obtain the required RV precision, and the median of the instrumental drift, monitored through the second fiber of HARPS-N, was 0.39 m s$^{-1}$, with a r.m.s. of the total time series of 0.76 m s$^{-1}$.
The spectra were obtained with 900s integration time and the average value of the signal-to-noise ratio (S/N) is 110 per pixel on the extracted spectrum at 5500 \AA.
The data reduction and the RV measurements were performed by means of the data reduction software of HARPS-N \citep{2002A&A...388..632P}. Through the method of the CCF of the acquired spectrum by using a mask that depicts spectral features of a G2 star, we obtained the RV measurements listed in Table \ref{RV}.
The median of the internal errors is $\sigma_{\rm RV}$ = 0.7 m s$^{-1}$.
\begin{longtab}
\begin{longtable}{ccccccccc}
\caption{\label{RV} Time series of the HARPS-N spectra for HD\,108874: radial velocity (RV), bisector span (BVS), $\log R'_{\rm HK}$ and H$\alpha$ indices with the corresponding uncertainties.}\\
\hline\hline
\noalign{\smallskip}
BJD$_{\rm UTC}$ - 2 450 000 & RV & RV$_{\rm Err}$ & BVS & BVS$_{\rm Err}$ & $\log R'_{\rm HK}$& $\log R'_{\rm HK\,Err}$& H$\alpha$ &  H$\alpha_{\rm Err}$\\
& [km s$^{-1}$] & [km s$^{-1}$] & [km s$^{-1}$]  & [km s$^{-1}$] & & & &   \\
\hline
\noalign{\smallskip}
\endfirsthead
\caption{continued.}\\
\hline
\noalign{\smallskip}
\endhead
\hline
\noalign{\smallskip}
\endfoot
\hline
\noalign{\smallskip}
6266.775614 & -30.0006 & 0.0006 & -0.0337 & 0.0011 & -5.0542 & 0.0050 & 0.1964 & 0.0003 \\
6288.760070 & -30.0026 & 0.0009 & -0.0353 & 0.0017 & -5.0497 & 0.0100 & 0.1955 & 0.0008 \\
6297.779734 & -30.0175 & 0.0014 & -0.0385 & 0.0027 & -4.9875 & 0.0185 & 0.1955 & 0.0007 \\
6298.746416 & -30.0140 & 0.0005 & -0.0336 & 0.0010 & -5.0605 & 0.0045 & 0.1959 & 0.0003 \\
6299.685758 & -30.0171 & 0.0004 & -0.0368 & 0.0009 & -5.0555 & 0.0035 & 0.1968 & 0.0003 \\
6305.775108 & -30.0195 & 0.0005 & -0.0345 & 0.0010 & -5.0649 & 0.0049 & 0.1961 & 0.0004 \\
6324.745051 & -30.0251 & 0.0012 & -0.0384 & 0.0024 & -5.0072 & 0.0159 & 0.1953 & 0.0007 \\
6324.813447 & -30.0249 & 0.0008 & -0.0316 & 0.0016 & -5.0202 & 0.0080 & 0.1995 & 0.0005 \\
6344.655959 & -30.0420 & 0.0004 & -0.0353 & 0.0009 & -5.0424 & 0.0036 & 0.1962 & 0.0003 \\
6345.557661 & -30.0451 & 0.0006 & -0.0321 & 0.0012 & -5.0386 & 0.0062 & 0.1964 & 0.0003 \\
6362.640907 & -30.0451 & 0.0004 & -0.0374 & 0.0009 & -5.0504 & 0.0038 & 0.1988 & 0.0003 \\
6363.649495 & -30.0442 & 0.0005 & -0.0363 & 0.0011 & -5.0549 & 0.0049 & 0.1960 & 0.0003 \\
6364.676554 & -30.0460 & 0.0005 & -0.0389 & 0.0011 & -5.0476 & 0.0048 & 0.1968 & 0.0003 \\
6365.679606 & -30.0477 & 0.0005 & -0.0355 & 0.0010 & -5.0347 & 0.0044 & 0.1966 & 0.0003 \\
6366.546255 & -30.0501 & 0.0011 & -0.0324 & 0.0022 & -5.0701 & 0.0150 & 0.1995 & 0.0007 \\
6375.557854 & -30.0540 & 0.0009 & -0.0383 & 0.0017 & -5.0353 & 0.0098 & 0.1967 & 0.0005 \\
6376.556491 & -30.0521 & 0.0005 & -0.0348 & 0.0010 & -5.0531 & 0.0042 & 0.1959 & 0.0003 \\
6379.619984 & -30.0550 & 0.0006 & -0.0346 & 0.0012 & -5.0501 & 0.0057 & 0.1960 & 0.0003 \\
6380.611045 & -30.0558 & 0.0009 & -0.0366 & 0.0017 & -5.0430 & 0.0109 & 0.1968 & 0.0005 \\
6382.656900 & -30.0546 & 0.0005 & -0.0369 & 0.0009 & -5.0488 & 0.0039 & 0.1983 & 0.0002 \\
6398.553192 & -30.0536 & 0.0007 & -0.0343 & 0.0015 & -5.0761 & 0.0085 & 0.1962 & 0.0005 \\
6399.504679 & -30.0494 & 0.0017 & -0.0369 & 0.0034 & -5.1361 & 0.0372 & 0.1996 & 0.0009 \\
6404.538609 & -30.0506 & 0.0007 & -0.0340 & 0.0015 & -5.0350 & 0.0073 & 0.1967 & 0.0004 \\
6407.620984 & -30.0504 & 0.0007 & -0.0329 & 0.0013 & -5.0365 & 0.0069 & 0.1987 & 0.0003 \\
6408.661413 & -30.0494 & 0.0009 & -0.0357 & 0.0018 & -5.0486 & 0.0118 & 0.1960 & 0.0004 \\
6418.483200 & -30.0496 & 0.0013 & -0.0314 & 0.0026 & -5.0992 & 0.0216 & 0.1982 & 0.0007 \\
6428.371051 & -30.0499 & 0.0005 & -0.0326 & 0.0009 & -5.0741 & 0.0039 & 0.1958 & 0.0003 \\
6430.401707 & -30.0475 & 0.0005 & -0.0344 & 0.0010 & -5.0863 & 0.0044 & 0.1957 & 0.0003 \\
6483.375117 & -30.0103 & 0.0010 & -0.0314 & 0.0020 & -5.0603 & 0.0129 & 0.1981 & 0.0007 \\
6483.378856 & -30.0106 & 0.0012 & -0.0328 & 0.0024 & -5.0714 & 0.0169 & 0.1993 & 0.0008 \\
6483.382709 & -30.0118 & 0.0012 & -0.0289 & 0.0024 & -5.0736 & 0.0175 & 0.1976 & 0.0008 \\
6616.763764 & -30.0084 & 0.0010 & -0.0363 & 0.0021 & -5.0226 & 0.0130 & 0.1958 & 0.0005 \\
6617.783186 & -30.0085 & 0.0006 & -0.0292 & 0.0013 & -5.0470 & 0.0065 & 0.1958 & 0.0003 \\
6618.749862 & -30.0094 & 0.0006 & -0.0359 & 0.0012 & -5.0502 & 0.0061 & 0.1969 & 0.0003 \\
6655.743085 & -30.0243 & 0.0024 & -0.0354 & 0.0049 & -4.8748 & 0.0359 & 0.1975 & 0.0013 \\
6693.806103 & -30.0448 & 0.0006 & -0.0349 & 0.0012 & -5.0367 & 0.0059 & 0.1958 & 0.0003 \\
6698.600386 & -30.0445 & 0.0007 & -0.0341 & 0.0013 & -5.0318 & 0.0066 & 0.1953 & 0.0003 \\
6728.754863 & -30.0602 & 0.0006 & -0.0327 & 0.0013 & -5.0500 & 0.0070 & 0.1958 & 0.0003 \\
6762.528858 & -30.0741 & 0.0006 & -0.0349 & 0.0011 & -5.0648 & 0.0056 & 0.1988 & 0.0004 \\
6763.510519 & -30.0760 & 0.0008 & -0.0351 & 0.0015 & -5.0624 & 0.0091 & 0.1961 & 0.0005 \\
6764.391997 & -30.0752 & 0.0019 & -0.0315 & 0.0038 & -5.0981 & 0.0408 & 0.1954 & 0.0010 \\
6768.504113 & -30.0763 & 0.0022 & -0.0466 & 0.0043 & -5.0081 & 0.0431 & 0.1974 & 0.0011 \\
6769.460455 & -30.0740 & 0.0006 & -0.0356 & 0.0012 & -5.0462 & 0.0060 & 0.1975 & 0.0003 \\
6775.416283 & -30.0777 & 0.0007 & -0.0302 & 0.0014 & -5.0650 & 0.0084 & 0.1944 & 0.0004 \\
6783.519267 & -30.0821 & 0.0005 & -0.0325 & 0.0010 & -5.0469 & 0.0046 & 0.1974 & 0.0003 \\
6784.454691 & -30.0830 & 0.0004 & -0.0331 & 0.0009 & -5.0525 & 0.0042 & 0.1980 & 0.0003 \\
6785.460630 & -30.0825 & 0.0008 & -0.0345 & 0.0015 & -5.0496 & 0.0092 & 0.1963 & 0.0005 \\
6786.544167 & -30.0853 & 0.0007 & -0.0292 & 0.0014 & -5.0458 & 0.0083 & 0.1978 & 0.0004 \\
6798.390881 & -30.0766 & 0.0004 & -0.0355 & 0.0009 & -5.0593 & 0.0039 & 0.1957 & 0.0003 \\
6799.439721 & -30.0773 & 0.0005 & -0.0367 & 0.0010 & -5.0459 & 0.0049 & 0.1958 & 0.0003 \\
6800.432767 & -30.0769 & 0.0005 & -0.0350 & 0.0010 & -5.0605 & 0.0050 & 0.1977 & 0.0003 \\
6801.391509 & -30.0766 & 0.0004 & -0.0335 & 0.0009 & -5.0616 & 0.0039 & 0.1975 & 0.0003 \\
6802.404274 & -30.0783 & 0.0007 & -0.0349 & 0.0015 & -5.0680 & 0.0090 & 0.1951 & 0.0004 \\
6803.408011 & -30.0776 & 0.0008 & -0.0321 & 0.0016 & -5.0565 & 0.0103 & 0.1947 & 0.0005 \\
6817.431312 & -30.0699 & 0.0004 & -0.0345 & 0.0009 & -5.0382 & 0.0040 & 0.1953 & 0.0003 \\
6818.451326 & -30.0681 & 0.0005 & -0.0330 & 0.0010 & -5.0311 & 0.0046 & 0.1952 & 0.0003 \\
6819.429828 & -30.0701 & 0.0005 & -0.0324 & 0.0011 & -5.0373 & 0.0050 & 0.1946 & 0.0003 \\
6819.502159 & -30.0690 & 0.0006 & -0.0351 & 0.0012 & -5.0209 & 0.0063 & 0.1970 & 0.0003 \\
6820.418328 & -30.0682 & 0.0006 & -0.0294 & 0.0012 & -5.0330 & 0.0061 & 0.1976 & 0.0003 \\
6821.466533 & -30.0672 & 0.0011 & -0.0263 & 0.0021 & -5.0539 & 0.0164 & 0.1974 & 0.0006 \\
6858.394236 & -30.0497 & 0.0007 & -0.0302 & 0.0014 & -5.0397 & 0.0083 & 0.1967 & 0.0003 \\
6859.400003 & -30.0494 & 0.0008 & -0.0308 & 0.0016 & -5.0472 & 0.0098 & 0.1973 & 0.0004 \\
6860.381757 & -30.0491 & 0.0005 & -0.0333 & 0.0010 & -5.0516 & 0.0050 & 0.1955 & 0.0003 \\
6986.762130 & -30.0201 & 0.0006 & -0.0345 & 0.0012 & -5.0519 & 0.0071 & 0.1958 & 0.0004 \\
7011.746428 & -30.0177 & 0.0005 & -0.0330 & 0.0010 & -5.0466 & 0.0046 & 0.1966 & 0.0003 \\
7027.636657 & -30.0290 & 0.0011 & -0.0293 & 0.0022 & -5.0626 & 0.0188 & 0.1954 & 0.0006 \\
7028.663761 & -30.0283 & 0.0007 & -0.0320 & 0.0013 & -5.0508 & 0.0078 & 0.1961 & 0.0004 \\
7069.717576 & -30.0439 & 0.0010 & -0.0318 & 0.0021 & -5.0433 & 0.0151 & 0.1966 & 0.0006 \\
7075.792096 & -30.0425 & 0.0009 & -0.0332 & 0.0018 & -5.0332 & 0.0125 & 0.1956 & 0.0005 \\
7095.461326 & -30.0531 & 0.0008 & -0.0334 & 0.0015 & -5.0522 & 0.0094 & 0.1981 & 0.0004 \\
7097.461977 & -30.0533 & 0.0007 & -0.0338 & 0.0015 & -5.0571 & 0.0096 & 0.1962 & 0.0005 \\
7099.486939 & -30.0570 & 0.0007 & -0.0372 & 0.0014 & -5.0522 & 0.0083 & 0.1953 & 0.0004 \\
7108.714635 & -30.0623 & 0.0011 & -0.0363 & 0.0023 & -5.0437 & 0.0183 & 0.1953 & 0.0006 \\
7117.454735 & -30.0603 & 0.0005 & -0.0334 & 0.0010 & -5.0515 & 0.0051 & 0.1950 & 0.0003 \\
7118.479650 & -30.0610 & 0.0006 & -0.0362 & 0.0013 & -5.0491 & 0.0074 & 0.1951 & 0.0004 \\
7119.455723 & -30.0686 & 0.0014 & -0.0334 & 0.0028 & -5.0352 & 0.0235 & 0.1963 & 0.0009 \\
7120.443969 & -30.0665 & 0.0011 & -0.0364 & 0.0021 & -5.0226 & 0.0156 & 0.1951 & 0.0005 \\
7123.523025 & -30.0605 & 0.0009 & -0.0318 & 0.0018 & -5.0628 & 0.0127 & 0.1970 & 0.0006 \\
7137.632750 & -30.0681 & 0.0006 & -0.0276 & 0.0012 & -5.0440 & 0.0065 & 0.1985 & 0.0003 \\
7139.580028 & -30.0758 & 0.0007 & -0.0325 & 0.0014 & -5.0586 & 0.0081 & 0.1955 & 0.0004 \\
7140.588987 & -30.0741 & 0.0008 & -0.0303 & 0.0016 & -5.0475 & 0.0103 & 0.1978 & 0.0004 \\
7148.463238 & -30.0808 & 0.0006 & -0.0334 & 0.0012 & -5.0732 & 0.0068 & 0.1962 & 0.0004 \\
7153.443246 & -30.0750 & 0.0007 & -0.0310 & 0.0014 & -5.0498 & 0.0078 & 0.1954 & 0.0004 \\
7154.460859 & -30.0770 & 0.0010 & -0.0347 & 0.0021 & -5.0649 & 0.0158 & 0.1953 & 0.0005 \\
7156.425289 & -30.0770 & 0.0010 & -0.0342 & 0.0019 & -5.0570 & 0.0132 & 0.1953 & 0.0004 \\
7177.491528 & -30.0810 & 0.0006 & -0.0344 & 0.0011 & -5.0695 & 0.0063 & 0.1975 & 0.0003 \\
7203.450604 & -30.0766 & 0.0009 & -0.0345 & 0.0019 & -5.0659 & 0.0144 & 0.1949 & 0.0005 \\
7204.410543 & -30.0767 & 0.0009 & -0.0319 & 0.0018 & -5.0563 & 0.0127 & 0.1940 & 0.0004 \\
7205.401614 & -30.0768 & 0.0005 & -0.0368 & 0.0010 & -5.0743 & 0.0049 & 0.1953 & 0.0003 \\
7207.394565 & -30.0793 & 0.0008 & -0.0332 & 0.0015 & -5.0887 & 0.0103 & 0.1969 & 0.0004 \\
7209.431218 & -30.0763 & 0.0008 & -0.0321 & 0.0017 & -5.0722 & 0.0125 & 0.1965 & 0.0005 \\
\end{longtable}
\end{longtab}

\section{Stellar parameters} \label{parameters}

The atmospheric parameters of HD\,108874 are measured as in \cite{2011A&A...525A..35B}, based on the line equivalent widths measurements, by using the 2013 version of the MOOG code \citep{1973ApJ...184..839S} and the line list in \cite{2012MNRAS.427.2905B}. The analysis is performed on a merged spectrum, obtained by coadding the available spectra of the target after the correction of the corresponding radial velocity shift, showing a S/N of $\sim$  1200 at 5500 \AA. A summary of the extracted parameters is presented in Table \ref{param}.
We used the web interface PARAM\footnote{\url{http://stev.oapd.inaf.it/cgi-bin/param}} \citep{2006A&A...458..609D} which is based on isochrones by \cite{Bressan2012} for the estimation of the stellar mass, radius, and age.
Besides the effective temperature and the metallicity, we also included the parallax (15.97$\pm$1.07 mas, \citealt{2007A&A...474..653V}) and the V mag of the star (8.76$\pm$0.02, \citealt{1997JApA...18..161Y}) as input quantities.
A general agreement with the literature is found for all of our estimates (e.g. \citealt{2005ApJS..159..141V}, Wi09, \citealt{2010A&ARv..18...67T}).
The spectral analysis yields a value of $v \sin i = 1.6\pm0.5$ km s$^{-1}$ (see \citealt{2011A&A...526A.103D}).
A calibration of the FWHM of the CCF,
using stars with transiting planets with known photometric rotation periods
observed with HARPS-N, provides a value of $v \sin i = 1.36\pm0.26$ km s$^{-1}$, which we adopt
in the following. 
The average value of the activity index $\log R'_{\rm HK}$, provided by the HARPS-N pipeline (see Sect. \ref{sec:rhk}), is equal to -5.05 and indicates that HD\,108874 is less active than the Sun. The rotation period measured in this work is also reported (see the following Sections).
\begin{table}
\caption[]{Stellar parameters of HD\,108874.}
\label{param}
$$
\begin{array}{p{0.5\linewidth}c}
\hline
\noalign{\smallskip}
Parameter & \mbox{Value} \\
\noalign{\smallskip}
\hline
\hline
\noalign{\smallskip}
{\it Extracted (this work):} & \\
T$_{\rm eff}$ (K) & 5585 \pm 20 \\
$\log g$ (cm s$^{-2}$) & 4.39 \pm 0.12 \\
$ [$FeI/H$] $ (dex) & 0.19 \pm 0.07 \\
$ [$FeII/H$] $ (dex) & 0.19 \pm 0.10 \\
Microturbolence (km s$^{-1}$) & 1.04 \pm 0.02 \\
$v \sin i$ (km s$^{-1}$) & 1.36\pm0.26 \\
\hline
\noalign{\smallskip}
{\it Estimated with PARAM:} & \\
Mass ($M_{\odot}$) & 0.996 \pm 0.032 \\
Radius ($R_{\odot}$) & 1.062 \pm 0.070 \\
Age (Gyr) & 6.48 \pm 3.47 \\
\hline
\noalign{\smallskip}
$\log R'_{\rm HK} $ & -5.05 \\
\noalign{\smallskip}
$P_{\rm rot}$ (d) & 40.20\pm 0.15  \\
\noalign{\smallskip}
\hline
\end{array}
$$
\end{table}

\section{Radial velocity analysis} \label{rva}
\begin{figure}
\centering
\includegraphics[width=0.7\linewidth,angle=-90]{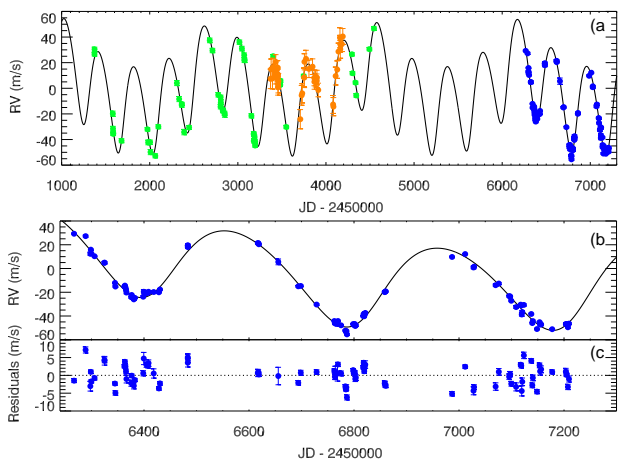}
\caption{{\it Panel a:} RV dataset for HD\,108874. Green dots: HIRES@Keck; orange dots: HRS@HET; blue dots: HARPS-N@TNG; black solid line: global fit of the three sets. {\it Panel b:} RV time series for HARPS-N data only. {\it Panel c:} Residuals after the subtraction of the two planets fit.
}
\label{totRV}
\end{figure}

\begin{figure}
\centering
\includegraphics[width=0.7\linewidth,angle=-90]{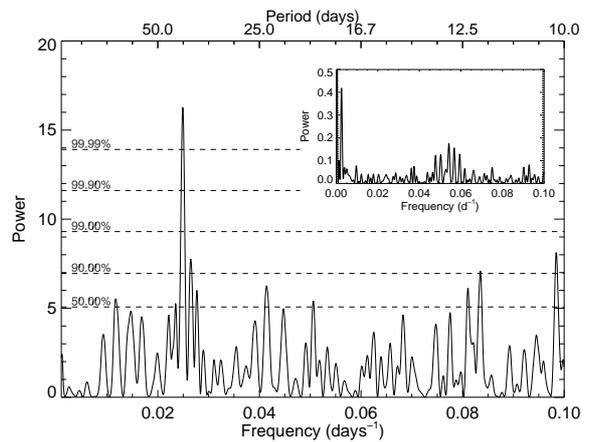}
\caption{Periodogram of RV residuals of HARPS-N dataset: both frequency and period domains are shown. Horizontal dashed lines represent the confidence level of the peaks. The inset shows the window function due to the temporal sampling of the data.
}
\label{freqP}
\end{figure}

\begin{figure}
\centering
\includegraphics[width=0.7\linewidth,angle=-90]{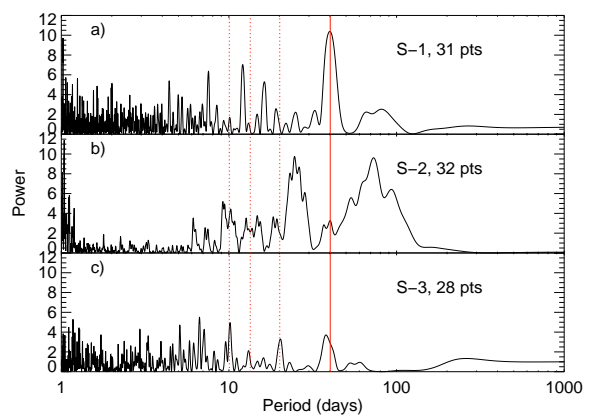}
\caption{Periodogram of HARPS-N RV residuals subdivided into each observing season. The red solid line marks the location of the 40.2 d periodicity, while the red dotted lines represent the 1$^{\rm st}$, 2$^{\rm nd}$ and the 3$^{\rm rd}$ harmonics of the main peak.
}
\label{seasons}
\end{figure}

We extended our RV dataset by considering the data available in the literature from HIRES@Keck and HRS@HET (see Tab. \ref{sum_dat}), obtaining a total time span of 16 yrs.
\begin{table}
\caption[]{Summary of the datasets.}
\label{sum_dat}
$$
\begin{array}{p{0.34\linewidth}cccc}
\hline
\noalign{\smallskip}
 \small{\mbox{Instrument}}   &\small{\mbox{Nr.}} & \small{\mbox{ Time span}} & \small{\overline{\mbox{RV}_{err}}} &\small{\mbox{Ref.}}\\
  &  \small{\mbox{epochs}}  & \small{\mbox{[d]}} & \small{[\mbox{m s}^{-1}]} &  \\
\noalign{\smallskip}
\hline
\hline
\noalign{\smallskip}
\small{HIRES (Keck-10m)}  & \small{55} & \small{3172} & \small{1.58} & \small{\mbox{Wr09}}\\
\small{HRS (HET-9.2m)}  & \small{40} & \small{820} & \small{6.53} & \small{\mbox{Wi09}} \\
\small{HARPS-N (TNG-3.6m)}  & \small{94} & \small{943} & \small{0.79}& \small{\mbox{-}} \\
\noalign{\smallskip}
\hline
\end{array}
$$
\end{table}
We performed a two-planets fit, including a zero-point correction of the RVs (see Sect. \ref{orbit}), obtaining the model overplotted to the three sets of data in panel $a$ of Fig. \ref{totRV}. Panel $b$ shows only HARPS-N data.
After the subtraction of this fit from the original time series, the resulting residuals show an r.m.s. of 4.1 m s$^{-1}$ for HIRES data, 5.9 m s$^{-1}$ for HRS (smaller than the typical errors) and 2.8 m s$^{-1}$ for HARPS-N. The r.m.s. of the residuals for the whole dataset is 4.03 m s$^{-1}$.
The Lomb-Scargle periodogram of the RV residuals of HARPS-N (Fig. \ref{freqP}) shows a clear periodicity at
$\sim$ 0.0248 d$^{-1}$ with an uncertainity of 1.0\,$\cdot 10^{-4}$ d$^{-1}$ evaluated with the relation in \cite{1999DSSN...13...28M}, corresponding to $40.20\pm\rm0.15$ d in the domain of periods, with a normalized power equal to 16.3 \citep{1986ApJ...302..757H} and a confidence level higher than 99.99\%, obtained after 100,000 bootstrap random permutations.
The spectral window of the HARPS-N dataset has been computed (inset of Fig. \ref{freqP}) as in \cite{1975Ap&SS..36..137D}: the first relevant peak in our region of interest is at 0.0025 d$^{-1}$, corresponding to one cycle per year and is due to the visibility of the star.
Many other small peaks are also present between 0.045 d$^{-1}$ and 0.065 d$^{-1}$ (i.e., 15-22 d),
probably due to the scheduling of the GAPS observing runs.
Therefore, it does not seem that the periodicity of 40.2 d is due to an aliasing effect with the orbital periods of the two planets.
Being this periodicity very robust, we performed a three-planets fit to the data, with the addition of a further Keplerian function aiming to model the signal in the RV residuals. The resulting r.m.s of the time series decreased from 4.03 to 2.2 m s$^{-1}$ and the periodogram does not show any residual power around 40.2 d.
The fit implies a minimum mass of about 17 M$_{\oplus}$ and a circular orbit for the planet candidate. 
Few not significant peaks are found around 40 d in the periodogram of residuals for HIRES and HRS, explainable by sub-optimal temporal sampling, but there is also the possibility that this signal is not driven by Keplerian motion. For this reason we analysed the single observing seasons of HARPS-N in order to verify the presence of this periodicity from one year to the next (Fig. \ref{seasons}). In the figure, the location of $P=40.2$ d is indicated as a reference, and the second and the third harmonics of the main peak are also shown (20.1 d, 13.4 d and 10.05 d). 
The signal at 40.2 d is present only in the first season (panel $a$, with a confidence level of 99.97\%), and marginally in the third one (panel $c$), together with its harmonics.
In the second season (panel $b$) the 40.2 d peak is suppressed: we only observe two large features at 25 and 73 d probably associated to each other. 


\section{Stellar activity} \label{act}

\subsection{Log $R'_{\rm HK}$ and H$\alpha$} \label{sec:rhk}
\begin{figure}
\centering
\includegraphics[width=0.7\linewidth,angle=-90]{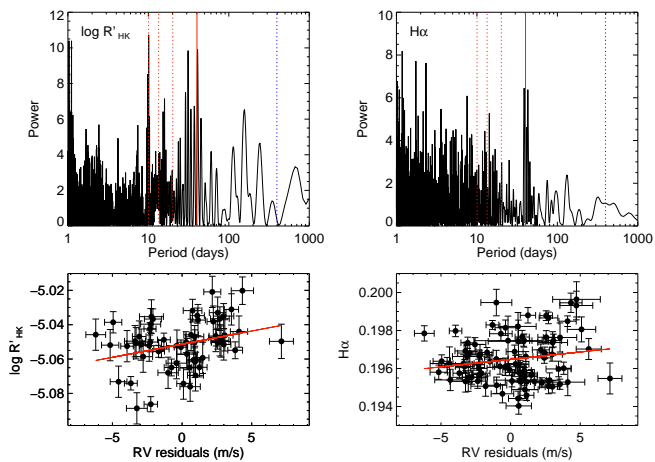}
\caption{{\it Upper panels}: Periodograms of the $\log R'_{\rm HK}$ (with $S/N>20$) and of the H$\alpha$ indices for HARPS-N data. The red solid line indicates the 40.2 d period, the red dotted lines represent its 1$^{\rm st}$, 2$^{\rm nd}$ and the 3$^{\rm rd}$ harmonics; the blue dotted line indicates the period of planet b. {\it Lower panels}: Correlation with the RV residuals.
}
\label{rhk}
\end{figure}

The chromospheric emission from Ca\textrm{II} H and K lines of HARPS-N spectra (log $R'_{\rm HK}$) is directly provided by the HARPS-N pipeline \citep{2011arXiv1107.5325L}. 
We added the S-index measurements obtained from the HIRES dataset by \cite{2004ApJS..152..261W} (both datasets are calibrated with the Mt. Wilson activity survey, so no offset should be present) and converted into log $R'_{\rm HK}$ by using the scaling relations by \cite{1984ApJ...279..763N}.
We perform a tentative sinusoidal fit of the stellar activity cycle with the Levenberg–Marquardt fitting algorithm through the IDL package {\tt MPFIT}\footnote{\tiny{\url{https://www.physics.wisc.edu/~craigm/idl/fitting.html}}} (Fig. \ref{activity}, {\it a}). According to the parameters of our fit, the length of this cycle is $\sim$19 years, but the huge gap in the data between 2003 and 2012 and an insufficient temporal coverage does not ensure the goodness of the fit. 
Since the long term correction does not lead to different results, we show the results only for the uncorrected time series.
The upper left panel of Fig. \ref{rhk} shows the periodogram of the $\log R'_{\rm HK}$. The periodicity at 40.2 d detected in the periodogram of the RV residuals is shown, located very close to one of the strongest peaks in the plot, having a statistical significance of 99.75\%. Its third harmonics could be the responsible for the largest periodicity in the periodogram, around 10 d. In the lower left panel we compare the RV residuals with the values of $\log R'_{\rm HK}$: they show a moderate positive correlation with Spearman and Pearson coefficients of $\sim 0.3$ (see Table \ref{tbl:asy} for a summary of correlation coefficients), slightly lower than the values obtained by \cite{2016A&A...587A.103L} in the solar case, 0.35 and 0.38, respectively.
This is a remarkable indication of the physical origin of the 40.2 d periodicity, leading us to identify it with the rotation period, $P_{\rm rot}$, of the star. Furthermore, following \cite{2008ApJ...687.1264M} the expected rotation period for our target is 39.9 d, derived from the mean value of $\log R'_{\rm HK}$ and the $B-V$ (0.738).
This value, considering the stellar radius derived in Sect. \ref{parameters}, is also consistent with the adopted $v \sin i$ for an edge-on inclination: larger $v \sin i$ yields unphysical values for the inclination.
We also analysed the periodograms of the $\log R'_{\rm HK}$ measurements for separate seasons (Table \ref{tbl:asy}). As in the case of the RVs residuals, the periodicity at 40.2 d is clearly present only in the first season while we only observe a periodicity of $\sim$20 d (first harmonic of $P_{\rm rot}$) in the third one, also present in the RV residuals.
 
The HARPS-N spectra of HD\,108874 were also analysed to extract the time series of the H$\alpha$ index (following \citealt{2011A&A...534A..30G}): the periodogram analysis reveals an excess of power distributed in a narrow envelope of peaks around 40 d, in agreement with the adopted $P_{\rm rot}$ (upper right panel of Fig. \ref{rhk}). No significant correlation is found with the RV residuals (see Tab. \ref{tbl:asy}).
Despite the periodograms of the H$\alpha$ and $\log R'_{\rm HK}$ indices show similar periodicities, a very weak correlation is found between these two quantities ($C_{\rm P}=0.18$, $p$-value=0.16). This can be explained by the presence of plages on the stellar disc, revealed both by the Ca II and H$\alpha$ lines, and filaments able to modify the emission of the H$\alpha$ and to break the correlation between the time series, as demonstrated by \cite{2009A&A...501.1103M}, \cite{2014A&A...566A..66G}, and by \cite{2016arXiv161005923S}, for early-M dwarfs. When the different timescales of these phenomena are modulated by the stellar rotation, their periodograms can actually show periodicities related to it.

\subsection{Asymmetry indicators of the CCF} \label{sec:asy}
\begin{figure}
\centering
\includegraphics[width=0.6\linewidth,angle=-90]{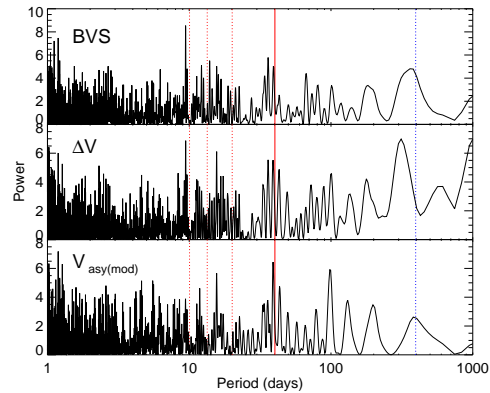}
\caption{Periodograms of the asymmetry indicators as derived by the HARPS-N pipeline (BVS) and Lanza et al. (in prep; $\Delta V$, $V_{\rm asy(mod)}$). Red solid and dotted lines represent the assumed $P_{\rm rot}$ and its 1$^{\rm st}$, 2$^{\rm nd}$ and the 3$^{\rm rd}$ harmonics, respectively, while the blue dotted line indicates the period of planet b.
}
\label{asy}
\end{figure}

Stellar activity also results in a deformation of the line profile of the spectral lines, which can be quantified by several asymmetry indicators. We investigated these quantities that provide an independent evaluation of the stellar activity with respect to the chromospheric indices.
A measurement of the CCF bisector velocity span (BVS) is directly provided by the HARPS-N pipeline, while a procedure presented in Lanza et al. (in prep.) estimates the values and the errors of $\Delta V$ (representing the RV shift produced by the asymmetry alone as defined in \citealt{2006A&A...453..309N} and reconsidered by \citealt{2013A&A...557A..93F}) and of the quantity $V_{\rm asy(mod)}$\footnote{\tiny{\url{https://sites.google.com/a/yale.edu/eprv-posters/home}}}, a modified version of the $V_{\rm asy}$ as defined by \cite{2013A&A...557A..93F} but not affected by the RV shifts as in the case of the original definition.
Periodograms of the asymmetry indices time series are presented in Fig. \ref{asy}: all of them show a moderate amount of power around 40 d.
Even in this case, there is no evidence of a clear correlation between the RV residuals and line profile indicators (Table \ref{tbl:asy}).

\begin{figure}
\centering
\includegraphics[width=5cm,angle=-90]{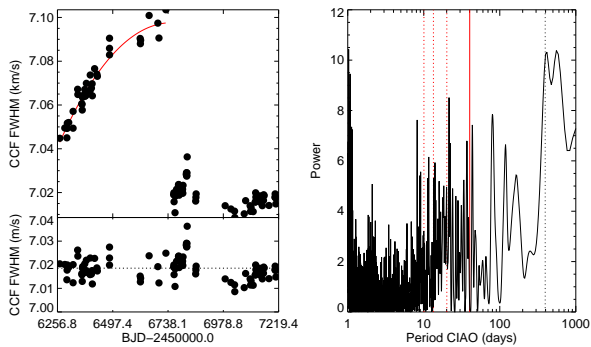}
\caption{{\it Left upper panel}: original time series of the CCF FWHM, showing the trend due to the defocusing. {\it Left lower panel}: residuals of the CCF FWHM after the subtraction of the fit. {\it Right panel}: Periodogram of the residual CCF FWHM. The red solid line represents the $P_{\rm rot}$, the dotted lines represent its 1$^{\rm st}$, 2$^{\rm nd}$ and the 3$^{\rm rd}$ harmonics. 
}
\label{fwhm}
\end{figure}

Finally we analysed the FWHM of the CCF, initially affected by the HARPS-N defocusing which progressively enlarged it (left upper panel of Fig. \ref{fwhm}).
We tried to remove this effect by performing a polynomial fit, shown in the figure, before the focus correction. The resulting residuals (left lower panel) produce the periodogram in the right panel of Fig. \ref{fwhm}. Besides a long-term periodicity close to the period of planet b, we found a peak around 20 d ($P_{\rm rot}/2$) and some power close to $P_{\rm rot}$, indicated on the figure.
The correlation coefficients between RV residuals and the corrected FWHM in Table \ref{tbl:asy} (C$_{\rm P} \sim \rho \sim 0.2$), show a very weak linear correlation.

\begin{table}
\caption[]{Summary of the values of Pearson (C$_{\rm P}$) and Spearman ($\rho$) correlation coefficients and corresponding $p$-values between RV residuals and all the activity indices investigated in this work.
}
\label{tbl:asy}
$$
\begin{array}{p{0.3\linewidth}cccc}
\hline
\noalign{\smallskip}
& \mbox{C}_{\rm P} & p\mbox{-value} & \rho & p\mbox{-value}\\
\hline
\hline
\noalign{\smallskip}
$\log R'_{\rm HK}$ & 0.30 & 0.02 & 0.32 & 0.01 \\
$\log R'_{\rm HK}$ (season 1) & 0.30 & 0.18 & 0.28 & 0.21 \\
$\log R'_{\rm HK}$ (season 2) & 0.30 & 0.06 & 0.35 & 0.03 \\
$\log R'_{\rm HK}$ (season 3) & 0.44 & 0.10 & 0.55 & 0.04 \\
H${\alpha}$ & 0.13 & 0.20 & 0.06 & 0.55 \\
BVS & -0.01 & 0.60 & -0.02 & 0.80 \\
$\Delta V$ & -0.03 & 0.75 & -0.09 & 0.41 \\
$V_{\rm asy(mod)}$ & 0.08 & 0.47 & 0.14 & 0.17 \\
FWHM$_{\rm CCF}$ & 0.22 & 0.04 & 0.17 & 0.11 \\
\noalign{\smallskip}
\hline
\end{array}
$$
\end{table}

\subsection{Activity modelling of raw time series} \label{model}
\begin{figure}
   \centering
  \includegraphics[width=5cm,angle=-90]{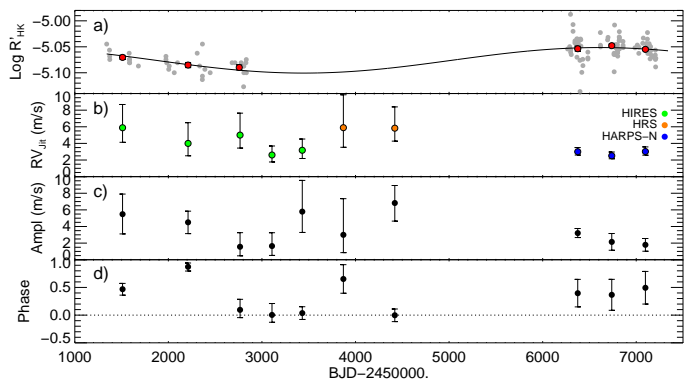}
      \caption{{\it a)}: Tentative fit of the stellar activity cycle for HARPS-N and HIRES data of log $R'_{\rm HK}$. Data points (grey dots) are plotted together with the mean values of each observing season (red dots). $b)$: fitted RV jitters for the separate observing seasons of HIRES, HRS and HARPS-N datasets, when activity model is not included in the data; $c)$: semi-amplitude of the fitted RV sinusoid; $d)$: phase of the fitted RV sinusoid.
            }
         \label{activity}
   \end{figure}

We modelled the contribution of the activity in the full RV dataset from HIRES, HRS and HARPS-N, following the approach of \cite{2011A&A...528A...4B} as recently implemented in the open-source code \texttt{PyORBIT}\footnote{Available at \url{https://github.com/LucaMalavolta/PyORBIT}} (see \citealt{2016A&A...588A.118M} for the details on the activity model, its implementation and the steps involved in parameter estimation).
Due to the lack of simultaneous photometric data, we used the value of $P_{\rm rot}$ from the activity indexes as prior and constrained it within $  \pm 0.5$ d.
We considered several combinations of harmonics for each dataset, and in all cases we found that the harmonics had RV amplitudes consistent with zero, so we decided to use only the sinusoids associated to $P_{\rm rot}$ in the final fit.
When including the activity model in the RV fit, the jitter parameter is reduced by 30\% for the HIRES and HARPS-N datasets, while no improvement is visible in the (noisier) HRS dataset. The orbital parameters of the two planet companions are not affected by the activity correction since the RV modulation has a shorter time-scale with respect to their orbital periods, so the activity noise is averaged out during the fitting process.
In Fig. \ref{activity} (panels $b$, $c$, $d$) we show the behaviour of the activity with time by comparing the fit of the activity cycle of the star described in Sect. \ref{sec:rhk} ($a$), the seasonal values of the RV jitter parameter when the activity model is not included in the fit ($b$), and the semi-amplitude and phase of the fitted RV sinusoid respectively ($c$, $d$). The jitter terms of the HIRES dataset show a gradual reduction, which may explain the negative slope of the corresponding activity cycle on the left side of panel $a$. The HRS jitter is probably dominated by  instrumental errors, while HARPS-N shows a small amount of jitter, despite the correspondence at the maximum of the activity cycle. The amplitude of the sinusoid in the first season of HARPS-N data ($c$) is larger than that in the subsequent two seasons, explaining the excess of scatter in the $\log R'_{\rm HK}$ time series ($a$).
\\

\section{Refined orbital solution} \label{orbit}

To refine the orbital parameters of the two known planets around HD\,108874,
we modelled both the literature and our new HARPS-N data
with two Keplerians, by including three RV zero points and
three uncorrelated RV jitter terms for each dataset (HIRES, HRS, and HARPS-N).
The jitter terms that account for possible RV scatter exceeding the nominal error bars were added in quadrature to the RV uncertainties,
regardless of the origin of the jitter (stellar activity, instrumental effects, and/or a combination thereof).
In addition to the offset and jitter terms,
we fitted for the inferior conjunction times,
the orbital periods, the RV semi-amplitudes,  the orbital eccentricity, the argument
of periastron and $\sqrt{e}\sin{\omega}$ and $\sqrt{e}\cos{\omega}$ \citep{Ford2005}
of the two planets. Therefore, our model has 16 free parameters in total.
\begin{table}
\caption[]{Orbital parameters of HD\,108874 b and c as derived from the present analysis: periastron time ($T_{\rm P}$, derived from the fit), inferior conjunction time ($T_{\rm C}$, measured through the fitting process), orbital period ($P$), $\sqrt{e} \sin{\omega}$, $\sqrt{e} \cos{\omega}$, orbital eccentricity ($e$), argument of periastron ($\omega$), RV semi-amplitude ($K$) and the HARPS-N zero point correction ($\gamma_{\rm HARPS-N}$). The minimum planetary mass ($M_{\rm p} \sin{i}$) and the semi-major axis ($a$) are also derived. }
\label{fit}
$$
\begin{array}{p{0.3\linewidth}cc}
\hline
\noalign{\smallskip}
Parameter & \mbox{HD\,108874 b} & \mbox{HD\,108874 c}  \\
\noalign{\smallskip}
\hline
\hline
\noalign{\smallskip}
\noalign{\smallskip}
$T_{\rm C}$ $[$BJD$_{\rm TDB}]$ & 2454317.4 \pm 2.2 & 2454782.5 \pm 15.8  \\
\noalign{\smallskip}
$T_{\rm P}$ $[$BJD$_{\rm TDB}]$ &2454443.9 \pm 6.5  & 2454521.7 \pm 29.8 \\
\noalign{\smallskip}
$P\, [$d$]$ & 395.34 \pm 0.19 & 1732.2 \pm 9.8 \\
\noalign{\smallskip}
$\sqrt{e} \sin{\omega}$ & -0.23 \pm 0.03 & 0.09 \pm 0.05  \\
\noalign{\smallskip}
$\sqrt{e} \cos{\omega}$ & -0.29 \pm 0.03  & 0.47 \pm 0.04  \\
\noalign{\smallskip}
$e$ & 0.142 \pm 0.011  & 0.229 \pm 0.032 \\
\noalign{\smallskip}
$\omega$ $[$deg$]$ & 218.7 \pm 6.0 & 11.8 \pm 7.5 \\
\noalign{\smallskip}
$K\,[$m s$^{-1}]$ & 35.18 \pm 0.64 &  19.06 \pm 0.63  \\
\noalign{\smallskip}
$M_{\rm p} \sin{i}$ $[$M$_{\rm J}]$ & 1.25 \pm 0.10 &  1.09 \pm 0.16  \\
\noalign{\smallskip}
$a\, [$AU$]$ & 1.05 \pm 0.02 &  2.81 \pm 0.06  \\
\noalign{\smallskip}
$\gamma_{\rm HARPS-N}$ [km s$^{-1}$] & \multicolumn{2}{c}{ -30.0294 \pm 0.0008 }\\ [0.5pt] 
\noalign{\smallskip}
\hline
\end{array}
$$
\end{table}
We determined the posterior distributions of the model parameters with a Bayesian
differential evolution Markov chain Monte Carlo (DE-MCMC) approach
\citep{TerBraak2006, Eastmanetal2013}. We ran 32 chains and used the same criteria
as in \citet{Bonomoetal2014} and \citet{2014A&A...567L...6D} for the removal
of burn-in steps, convergence and good mixing of the chains.
For all the parameters we considered uninformative priors.
The values of fitted and derived system parameters and their $1\sigma$ uncertainties,
which were computed as the medians and the 15.86\% and 84.14\% quantiles of their posterior distributions,
are listed in Table \ref{fit}.
A general agreement is found between the literature and our results for the orbital parameters, except for the eccentricity of planet b. Our estimate, equal to $0.142\pm 0.011$, is close to the value reported by Wr09, confirming that the planet is slightly more eccentric than the value derived by Wi09 ($0.082 \pm 0.021$). A significant difference has been found for the orbital period of planet c: HARPS-N data revealed a larger value of the period that differs by 4$\sigma$ from the value found by Wi09 (P$_{\rm c} = 1620 \pm 24$ d), and by 2$\sigma$ from the one in Wr09 (P$_{\rm c} = 1680 \pm 24$ d). Our estimates rule out the supposed mean motion resonance (MMR) of 4:1, revealing that the system is close to a 9:2 MMR. 
Radial velocity residuals of the two-planets fit show no evidence of a long term trend, in agreement with the results presented in \cite{2016ApJ...821...89B}.
The minimum masses of HD\,108874 b and c are 1.255 and 1.094 $M_{\rm J}$, with semi-major axes of 1.05 and 2.81 AU, respectively.


\section{System stability} \label{stab}
\begin{figure}
\centering
\includegraphics[width=6.5cm,angle=-90]{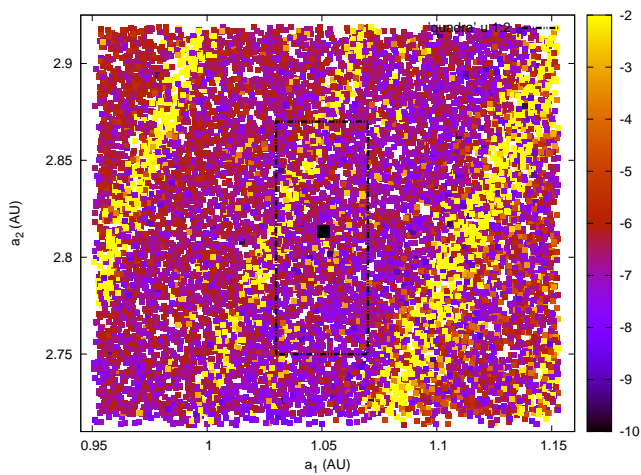}
\caption{Stability of the HD\,108874 system. The values of the semi-major axes of the two planets
are sampled around the nominal system (black square) according to the dispersion of the main secular frequency.
The black box marks the 1$\sigma$ uncertainties given in Table \ref{fit}.
The three yellow strips indicate the instability zones related to the 5:1, 9:2 and 4:1 MMR, respectively. A logarithm scale is used for the
colour coding, where small values of the dispersion mean stability, while large values imply fast changes of the secular frequency and then chaotic evolution.
}
\label{mmr}
\end{figure}

\begin{figure*}
\centering
\includegraphics[width=6cm,angle=-90]{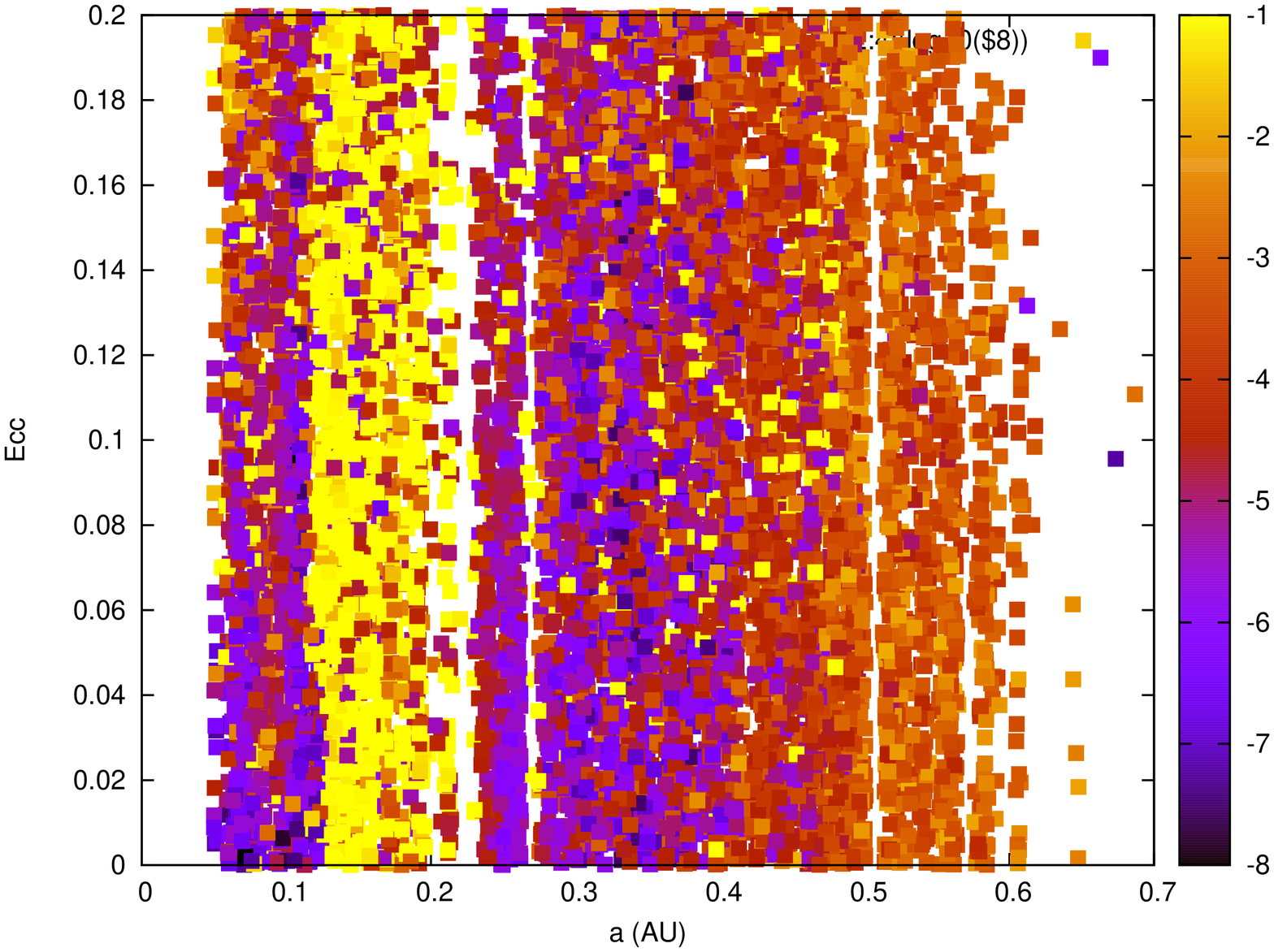}%
\includegraphics[width=6cm,angle=-90]{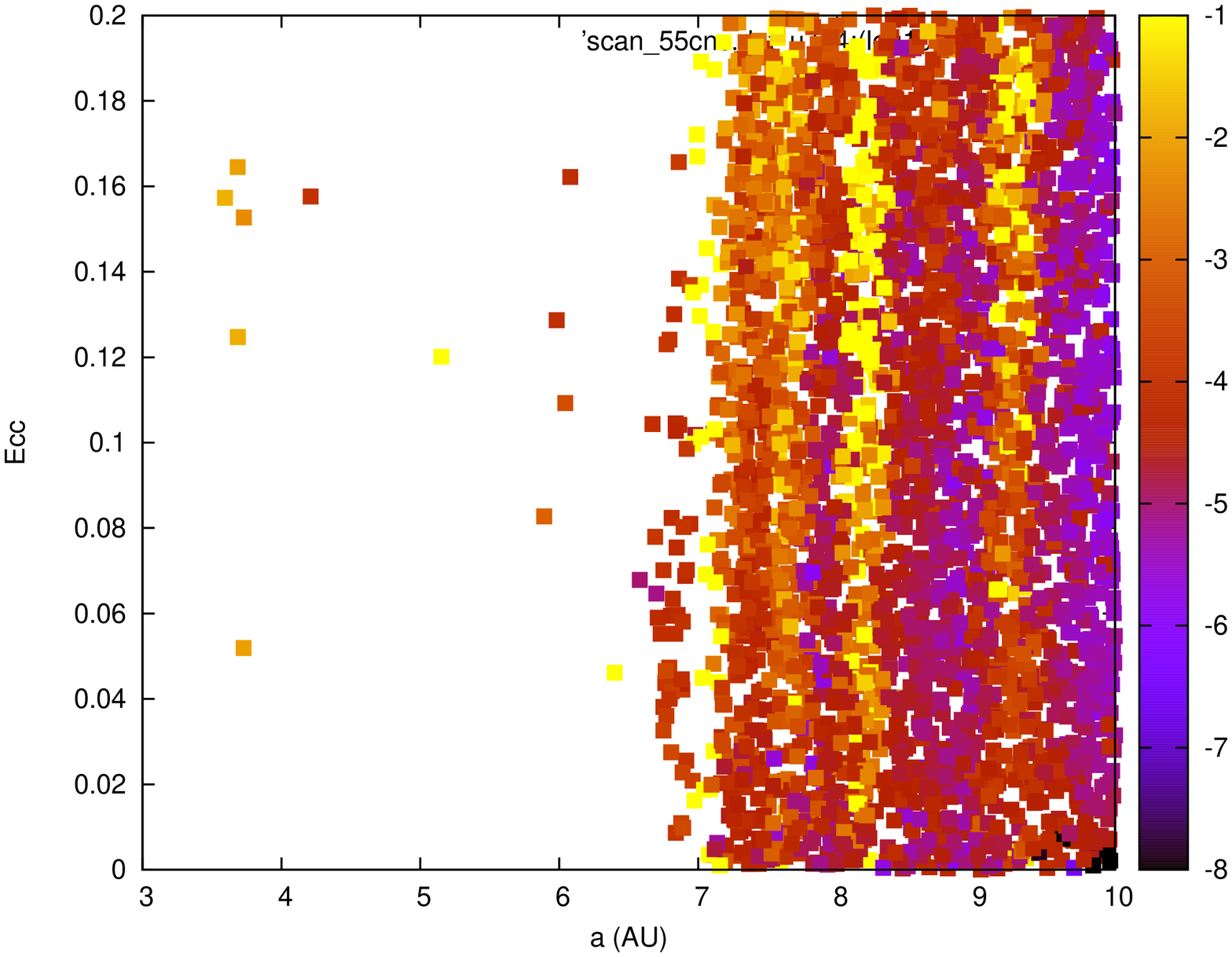}
\caption{Stability of a 1 M$_{\oplus}$ planet in the inner ({\it left}) and the outer regions ({\it right}) of the HD\,108874 system as a function of the initial  eccentricity and semi-major axis of the putative bodies. Colour scale ranges from black, representing stable regions, to yellow, which indicates higher degree of instability for planets with given separation and eccentricity. The empty (white) spaces indicate systems unstable over short timescales. }
\label{inout}%
\end{figure*}

We have investigated the long term stability of the system (by using the parameters in Table \ref{fit}) first with a long term numerical integration of the planet orbits, obtaining a `nominal' solution, and then with a parametric exploration of the phase space
around this solution. The direct N--body integration of the system over 1 Gyr, performed with SyMBA (the Symplectic Massive Body Algorithm; \citealt{1998AJ....116.2067D}), confirms that it is stable with a quasi-periodic behaviour over that timescale.
The exploration of the phase space around the nominal solution is performed with the Frequency Map Analysis (FMA; \citealt{lask93, sine96, marz03}). This is a numerical tool for the detection of chaotic behaviour based on the analysis 
of the variation of the secular frequencies of the system. Its main advantage is in allowing users
to identify unstable orbits with short term numerical integrations.
The orbits of about 40.000 systems
with orbital elements close to the nominal ones have been integrated with SyMBA over 5 Myr and their secular frequencies are analysed with the FMA to measure the diffusion speed in the phase space. Figure \ref{mmr} shows the outcome of the FMA as a function of the initial values of the semi-major axes ($a_1$ and $a_2$) of the two planets. The initial eccentricities have been randomly selected around the nominal values, the semi-major axes are sampled in an interval given by the nominal value $\pm$0.1 AU, while the initial inclinations are chosen between $0^{\circ}$ and $5^{\circ}$. The unknown angles, that is, the initial mean anomalies and node longitudes, have been randomly chosen between $0^{\circ}$ and $360^{\circ}$, while the pericentre argument is taken in between the nominal value $\pm$ $20^{\circ}$. 
The diffusion speed is measured as the dispersion of the main secular frequency of the system over running windows covering the integration timespan. The three large instability regions in Fig. \ref{mmr} are related to the 5:1, 9:2 and 4:1 mean motion resonances from left to right, respectively.
Our nominal system (shown in the figure, along with the current uncertainty limits) lies in a stable zone located close to the 9:2 resonance.
Close to a resonance, orbits may be chaotic while two planets trapped in MMR can be stable for long times depending on their location within the resonant region. A detailed
exploration of the resonant behaviour would be needed if the planets were in MMR, 
however in our case the resonance 
is present only in the outer margin of the uncertainty box so that, statistically, this configuration is 
less probable than a non--resonant one. 
As in Wi09, we  tested the possible existence of lower mass additional planets in the system. We performed a detailed investigation of the phase space between 0.5 and 10 AU within a full four-body model with the star, the two planets on their nominal orbits and a putative 1 M$_{\oplus}$ terrestrial planet. 
In Fig. \ref{inout} (left panel) we show the FMA of the region inside the orbits of the two nominal planets. 
Unstable systems over short timescales cannot be analysed and are represented by empty regions in the plots. By assuming $10^{-4}$ as a reference value between stable and unstable behaviour, a value tested with a few long term 
numerical integrations over 1 Gyr, two stable regions can be identified, one close to 0.1 AU and one extending 
from 0.25 to about 0.4 AU. 
Mean motion and secular resonances lead to chaotic evolution between 0.1 and 0.25 AU and beyond about 0.45 AU. In the region outside the two giant planets (right panel of Fig. \ref{inout}), the stability is met around 8.5 AU and beyond, especially for low values of eccentricity.

\section{The planetary system of HD\,108874} \label{discuss}
\subsection{Detection limits to additional planetary companions} \label{detlim}

\begin{figure}
\centering
\includegraphics[width=5cm,angle=-90]{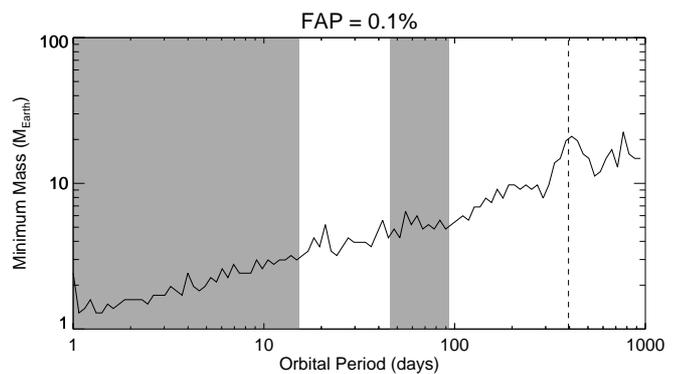}
\caption{Upper limits of the minimum mass of possible planetary companions to the HD\,108874 system, evaluated with a 99.9\% confidence level. Stable regions for an Earth mass planet are defined by grey areas; the vertical dashed line indicates the orbital period of planet b.
}
\label{lim}
\end{figure}

On the basis of the HARPS-N dataset we evaluated the upper limits of the minimum mass for possible planetary companions as in \cite{2009ApJ...697..544S}, with a 99\% confidence level, based on the F-test and $\chi^2$ statistics. 
Super Earths with $M\sin i \gtrsim 5$ M$_{\oplus}$ in the HD 108874 system interior of 0.5 AU 
 and objects with $M \sin i \gtrsim 2$ M$_{\oplus}$ with orbital periods of a few days could have been detected. 
The grey areas in Fig. \ref{lim} represent the allowed regions for a stable Earth mass planet, as derived in Sect. \ref{stab},
and the vertical dashed line indicates the orbital period of planet b.
The outcome of the dynamical analysis shows that no companions are allowed between the two giant planets, also excluded from HARPS-N data, at least for planets down to a few Neptune masses. The system is then dynamically full up to 7 AU, except for some stability strips in the inner region (Fig. \ref{inout}, left panel). As reported in Sect. \ref{stab}, a further planetary companion is potentially allowed in the outer region, beyond 11 AU: this hypothesis cannot be verified with the present instrumentation and the current observational baseline, at least for companions smaller than a brown dwarf. On the other hand, no evidence of a linear trend in the RV measurements is found (Sect. \ref{orbit} and \citealt{2016ApJ...821...89B}).

\subsection{Constraining the system architecture} \label{arch}
Our analysis indicates that, if present, planets with minimum mass down to 10 M$_{\oplus}$ should have been identified in the range between 1 and 200 d of period. The evidence from RV surveys (e.g. \citealt{2011arXiv1109.2497M}) and {\it Kepler} space-based photometry (e.g. \citealt{2013ApJS..204...24B}) points towards a high frequency of close-in, packed systems of super Earths and Neptunes, but only in systems without any detected outer giant companion.
Actually, from recent calculations \citep{2015ApJ...800L..22I} the populations of close-in super-Earth systems and Jupiter-like planets should be anti-correlated. Even if systems with hot Neptunes or hot super Earths with outer giant planetary companions do exist \citep{2016A&A...592A..13S}, systems hosting multiple super-Earths inside the orbits of giant planets are still uncovered. 
This configuration could be explained in terms of the role that giant planets may play in the terrestrial planet formation and evolution (e.g. \citealt{2003AJ....125.2692L}). \cite{2013ApJ...767..129M} performed simulations to investigate the fate of low-mass planets co-existing in the same planetary system with massive companions. Following their analysis, the architecture of the HD108874 system could be the result of a quiet dynamical evolution of the two giant planets, which leaves no modifications to their original eccentricities ($e<0.3$ for both of them and consistent with a planet migration mechanism, as confirmed in a study by  \citealt{2012MNRAS.427L..21R}) and orbital radii, but powerful enough to remove possible terrestrial planets through the ejection triggered by the secular perturbation or the merging with the star for crossing orbits with the inner giant planet. 
\cite{2015MNRAS.449L..65J} studied the relation between the minimum mass and the period ratio for adjacent giant planet pairs observed in multiplanet systems, obtaining a clear correlation which is also confirmed for the two planets hosted by HD\,108874. Their period--mass ratio falls well inside the level of scattering obtained with the calibrator pairs.
The period ratio of the analysed system ($\sim$4.4) is expected to be typical for pairs of giant planets around not evolved low-mass stars (i.e. $M_{\star} < 1.4 \, M_{\odot}$ and $\log g > 4.0$) when it is compared with the other systems considered, for example, in Fig. 1 in \cite{2016ApJ...819...59S}. In their analysis they consider also the total minimum mass of the planet system: for HD\,108874 it is 2.34$\rm\pm$0.26 $M_{\rm Jup}$, which appears to be larger with respect to the typical mass distribution.

\subsection{The habitable zone of HD\,108874} \label{hab}
We evaluated the limits of the habitable zone (HZ) of our target by exploiting the analytical relation by \cite{2007A&A...476.1373S}. Our results, reported in Table \ref{Tab:HZkp7}, are estimated for the classical early Mars and recent Venus criteria described in \cite{2013ApJ...765..131K} for the theoretical inner (runaway greenhouse) and outer limits with 50\% cloudiness (with H$_2$O and CO$_2$ clouds respectively) and the extreme theoretical limits, with a 100\% cloud cover.
\begin{table}
\caption{Boundaries of the habitable zone of HD\,108874.}
\label{Tab:HZkp7}
\begin{center}
\begin{tabular}{ c c c c c c c }
\hline
\noalign{\smallskip}
\multicolumn{7}{c}{Inner HZ [AU] ~~~~~~~~~~~~~~~~~~~~~~~~~~~ Outer HZ [AU]} \\
\cline{1-3}
\cline{5-7}
\noalign{\smallskip}
 \mbox{Venus} & \mbox{Clouds} & \mbox{Clouds} & & Mars & \mbox{Clouds} & \mbox{Clouds} \\
 & 50\% & 100\% & & &50\% & 100\% \\
\hline
\hline
\noalign{\smallskip}
 0.74 & 0.70& 0.47  & &  1.82  & 2.00  &2.46  \\
 \hline
\end{tabular}
\end{center}
\end{table}
Since HD\,108874 b is well inside the HZ we can consider the habitability of potential moons around it \citep{1997Natur.385..234W}.
Earth-mass moons revolving around Jupiter-mass planets have been shown to be dynamically stable for the lifetime of the solar system in systems where the stellar mass is larger than $0.15$ M$_\odot$ \citep{2002ApJ...575.1087B}. We considered this possibility in the case of HD\,108874 b: if the total mass of the planet-satellite system, $M_{\rm p} + M_{\rm s}$, satisfies the relation $M_{\rm s}\ll (M_{\rm p} + M_{\rm s}) \ll M_{\star}$, being $ M_{\star}$ the stellar mass, then $P_{\rm sat}\lesssim P_{\star\rm p}$, where $P_{\rm sat}$ is the satellite's orbital period and $P_{\star\rm p}$ is the circumstellar period of the planet-satellite system \citep{2012A&A...545L...8H}. This condition can be translated in an upper limit for the orbital radius of the satellite around the giant planet that we evaluated in a range of masses between Titan and ten Earth-masses.
\begin{figure}[htbp]
\centering
\includegraphics[width=6.5cm,angle=90]{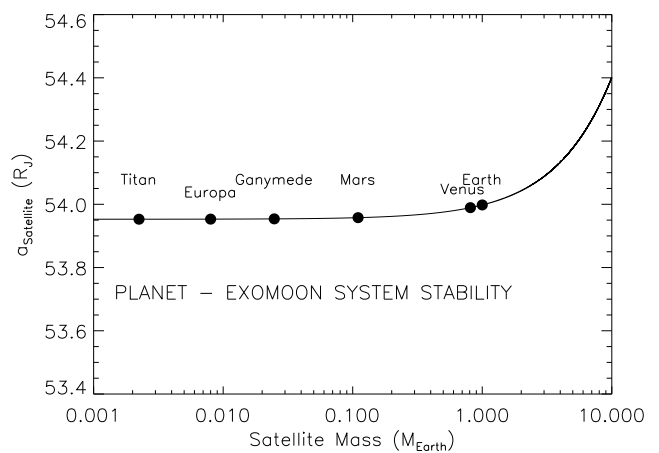}
\caption{Upper limit of the orbital separation between HD\,108874 b and an hypothetical satellite, evaluated for different values of the mass of the latter (smaller bodies of the solar system are depicted as reference).
}
\label{exom}
\end{figure}
The result is shown in Fig. \ref{exom}, where the position of a varying-mass satellite is identified with the name of the corresponding solar-system satellite or small planet and the orbital distance from the planet is indicated in Jupiter radii.
For masses up to one Earth mass the upper limit of the orbital radius is $\sim 53.95$ R$_{\rm J}$, about 0.025 AU.
In the case that HD\,108874 b transits its star the presence of the moon can be detected by measuring the variation in the transit time (TTV) of the planet due to gravitational effects, as shown by \cite{1999A&AS..134..553S}. According to their Eq. 24, if we consider satellites of the size of the Earth, the TTVs is approximately eleven minutes. The uncertainty of our ephemeris (a few days), obtained with RV data only is not suitable to detect such a signal, but in the case of transit observations, for example, with the forthcoming CHEOPS satellite, better constraints are expected.


\section{Conclusions} \label{concl}
We have presented the analysis of the intensive RV monitoring of the star hosting planets HD\,108874, with the HARPS-N spectrograph at TNG in the framework of the GAPS Programme.
A significant periodicity of 40.2 d has been found in the RV residuals of the two known planets fit but after a full analysis of the stellar activity we conclude that it must be addressed to the rotation period of the star and not the presence of an additional low-mass planet to the system.
This is an example of how the activity contribution to the RV is able to mimic a Keplerian modulation. 
We performed a refinement of the orbital parameters of the two giant planets, and the dynamical analysis shows that the system is in a stable configuration over 1 Gyr. 
Stable low-mass bodies are only allowed at small separation or very far from the star, even if HARPS-N data tend to exclude their presence.

\begin{acknowledgements}

This work was supported by 
INAF through the “Progetti Premiali” funding scheme of the Italian Ministry
of Education, University, and Research. 
The authors acknowledge Dr. L. Bedin (INAF-OAPD) and Dr. G. Lodato (Universit\`a degli Studi di Milano) for their comments and suggestions. We thank the anonymous referee for her/his useful advices and comments.
\end{acknowledgements}

\end{document}